\documentclass[toc]{PoS}

\usepackage{amsmath}
\usepackage{float}
\usepackage{mathtools}

\newcommand{\Tr}{\mathrm{Tr}}

\title{Open strings in integrable deformations of $\sigma$-models}

\ShortTitle{Open strings in integrable deformations of $\sigma$-models}

\author{\speaker{Sibylle Driezen}\\
Theoretische Natuurkunde, Vrije Universiteit Brussel  \& The International Solvay Institutes\\
Pleinlaan 2, B-1050 Brussels, Belgium,\\
and,\\
        Department of Physics, Swansea University\\Singleton Park, Swansea SA2 8PP, U.K.
	 \\
        E-mail: \email{sib.driezen@gmail.com}}

\abstract{
This contribution is based   on a talk given by the author at the ``Dualities and Generalized Geometries" session of the Corfu Summer Institute 2018 workshops.  We summarise the results of \cite{Driezen:2018glg}, focusing our attention on integrable $\lambda$-deformations of WZW models with boundaries.}

\FullConference{Corfu Summer Institute 2018 "School and Workshops on Elementary Particle Physics and Gravity"\\
		(CORFU2018)\\
		31 August - 28 September, 2018\\
		Corfu, Greece}

\begin{document}

\section{Introduction \label{Introduction}}

From the string worldsheet perspective, two-dimensional non-linear $\sigma$-models with boundaries provide a rich area  to describe curved background geometries with  D-brane configurations in string theory. These  non-perturbative degrees of freedom are essential higher-dimensional objects on which open strings can end, and of which the geometry is completely determined by the allowed worldsheet boundary conditions. The answer of what boundary conditions are \textit{allowed} is decided by symmetry. In the case of string theory, e.g., they should preserve worldsheet conformal invariance. For a $\sigma$-model describing  strings in curved backgrounds, the answer is usually challenging and tractable only when the precise (boundary) CFT description is available.\\
\indent A simple but non-trivial example where one can make progress is provided by the Wess-Zumino-Witten model \cite{Witten:1983ar} describing strings in group manifolds supported by an NS-flux. The exact conformal invariance of this model is  covered by the existence of two holomorphic currents underlying two copies of an affine Kac-Moody current algebra and two copies of a Virasoro algebra. The inclusion of boundaries in the WZW model has been studied in a number of works \cite{Kato:1996nu,Alekseev:1998mc,Felder:1999ka,Stanciu:1999id} by identifying maximally symmetric gluing conditions on the holomorphic currents at the boundary preserving one copy of both the Kac-Moody and Virasoro algebra. Although the former is not necessary for conformal invariance it leads to  an elegant geometrical picture of the allowed D-brane configurations: they should wrap \textit{twisted conjugacy classes} of the group manifold. For example in the $SU(2)_k$ WZW model one finds two  D0-branes and a further  $k-1$ D2-branes  that are blown up to wrap the conjugacy classes described by $S^2 \subset S^3$ \cite{Alekseev:1998mc}.\\
\indent When the precise CFT formulation is not available, however, we will see in this note an elegant D-brane picture can arise also in  the context of $\sigma$-models with worldsheet integrability. Integrable stringy $\sigma$-models attracted considerable attention since the observation of worldsheet integrability in the AdS$_5\times$S$^5$ superstring \cite{Bena:2003wd}. Classically,  they are characterised by the existence of an infinite number of local or non-local conserved charges in involution leading, in principle, to a dramatic simplicity and exact solvability. Including boundaries in the theory typically destroys  conserved charges such as, e.g., the loss of translational invariance through the boundary. In this note, we will focus on \textit{allowed}  boundary conditions that preserve  the classical integrable structure by demanding the conservation of a tower of non-local charges generated by a monodromy matrix. This method has been introduced in \cite{Cherednik:1985vs,Sklyanin:1988yz} and further developed from a classical string point of view in \cite{Dekel:2011ja}.\\
\indent A suitable integrable $\sigma$-model that makes contact between the above methods is the integrable $\lambda$-deformed WZW model introduced by Sfetsos in \cite{Sfetsos:2013wia}. The deformation parameter $\lambda \in [0,1]$ interpolates between the WZW model at $\lambda = 0$ and the non-Abelian T-dual of the Principal Chiral Model (PCM) in  a scaling limit $\lambda \rightarrow 1$. On ordinary Lie group manifolds, accommodating only bosonic field content, the deformation is marginally relevant \cite{Itsios:2014lca,Appadu:2015nfa}. However, significant evidence from both a worldsheet \cite{Hollowood:2014qma,Appadu:2015nfa} and target space \cite{Borsato:2016zcf,Chervonyi:2016ajp,Borsato:2016ose} perspective implies that, when applied to super-coset geometries, the $\lambda$-model is a truly marginal deformation introducing no Weyl anomaly. Hence, the deformation of the WZW group manifold can be thought of as a bosonic trunctation of a truly superstring theory. The question of establishing D-branes in this deformed geometry is therefore natural and has been pursued in  the article  \cite{Driezen:2018glg} on which this proceedings is based. We will see, by demanding integrability, that the geometrical picture of twisted conjugacy classes of the WZW point persists and naturally fits in the deformed geometry. The semi-classical flux quantisation will turn out to consistently be independent of the continuous $\lambda$-deformation parameter.   Additionally the $\lambda$-deformation allows to track the behaviour of D-branes under generalised dualities \cite{Driezen:2018glg} --again  a challenging question in general curved backgrounds-- by the non-Abelian T-dual scaling limit and Poisson-Lie T-duality to the integrable $\eta$-deformation of the PCM \cite{Vicedo:2015pna,Hoare:2015gda,Klimcik:2015gba}. Illustrated for the $G=SU(2)$ manifold one will find under both dualities D2-branes transforming  to space-filling D3-branes that can be shown to preserve the classical integrable structure of the dual theories.\\

We lay out in section \ref{s:bmm} the general procedure to construct \textit{integrable} boundary conditions of two-dimensional $\sigma$-models. We apply this method in section \ref{s:lambda} to the integrable $\lambda$-deformation where we first review the model's construction, then interpret the allowed integrable boundary conditions as twisted conjugacy classes (illustrated in the $G=SU(2)$ manifold) and discuss their behaviour under generalised T-dualities. We end with some conclusions and outlook directions in section \ref{s:concl}.

\section{The boundary monodromy method for integrable systems}\label{s:bmm}

The boundary monodromy method, introduced by Cherednik and Sklyanin in \cite{Cherednik:1985vs,Sklyanin:1988yz}, is a powerful tool to derive boundary conditions preserving the integrability property of two-dimensional integrable field theories. The method consists of demanding that a monodromy matrix constructed from a Lax connection generates an infinite tower of conserved \textit{non-local} charges when a  boundary is present\footnote{To have a truly classically integrable (boundary) theory one should moreover show these charges to Poisson commute. We will not discuss this here, but see e.g. \cite{Mann:2006rh}.}. We will briefly review it here, following \cite{Dekel:2011ja,Driezen:2018glg}, as well as the  case without boundaries  to introduce notations.

Let us first consider the no-boundary case in a general two-dimensional field theory on a periodic or infinite line. We denote the coordinates by $(\tau,\sigma)$  by analogy with the closed string worldsheet theory. It is known that an infinite tower of conserved charges can be generated when the equations of motion of the theory can be represented by a zero-curvature condition of a so-called $\mathfrak{g}^{\mathbb{C}}$-valued Lax connection ${\cal L}(z)$ that depends on a generic \textit{spectral} parameter $z \in \mathbb{C}$ \cite{Zakharov:1973pp},
\begin{equation} \label{eq:zerocurvatureLax}
\begin{aligned}
d {\cal L}(z) + {\cal L}(z) \wedge {\cal L}(z)  = 0 , \qquad \forall\, z \in \mathbb{C}.
\end{aligned}
\end{equation}
In this case the transport matrix  defined by,
\begin{equation}\label{eq:transport}
\begin{aligned}
T^\Omega(b,a ; z) =  \overleftarrow{P \exp} \left( - \int^b_a d\sigma\, \Omega [{\cal L}_\sigma (\tau,\sigma ;z ) ] \right) \in G^{\mathbb{C}},
\end{aligned}
\end{equation}
(with $\Omega : \mathfrak{g} \rightarrow \mathfrak{g}$  a constant Lie algebra automorphism included for generality) satisfies,
\begin{equation}\label{eq:TransportToTime}
\begin{aligned}
\partial_\tau T^\Omega (b,a ; z ) = T^\Omega (b,a ; z ) \Omega [{\cal L}_\tau (\tau , a ; z)] -  \Omega [{\cal L}_\tau (\tau , b ; z)] T^\Omega (b,a ; z ).
\end{aligned}
\end{equation}
Indeed, under periodic boundary conditions $\sigma \sim \sigma + 2\pi$ or asymptotic fall-off boundary conditions, one can then show that the monodromy matrix $T(2\pi,0 ; z)$ (for $\Omega = \mathbf{1}$) generates conserved charges by,
\begin{equation}
\begin{aligned}
\partial_\tau \Tr T(2\pi, 0 ; z )^n , \qquad \forall\ n\in \mathbb{N} \;\; \mathrm{ and } \;\; \forall\ z\in \mathbb{C} .
\end{aligned}
\end{equation}
Hence, every value of $n$ or every term in the  expansion of $\Tr T(2\pi, 0 ; z )$ in $z$, corresponds to a conserved charge.

When the two-dimensional theory is defined on a finite line $\sigma \in [0,\pi]$, describing by analogy an open string worldsheet theory, one can determine \textit{integrable} boundary conditions on the endpoints by demanding the production of conserved charges along similar lines as above. Reminiscent to the method of image charges, one can derive these by taking a  copy of the finite-line theory and act on it with a reflection $R: \sigma \rightarrow 2\pi - \sigma$. The boundary monodromy matrix $T_b (z)$ is then constructed by gluing the usual transport matrix $T(\pi, 0 ;z )$ in the original region to the transport matrix $T_R^\Omega (2\pi, \pi ; z)$ in the reflected region,
\begin{equation}
T_b (\lambda) = T_R^\Omega (2\pi, \pi ;z) T(\pi, 0 ;z).
\end{equation}
Notice that in the reflected region we have included the possibility of a non-trivial automorphism $\Omega$ acting on the Lax connection in the path-ordered exponential as in \eqref{eq:transport}. By demanding that  the time derivative of the boundary monodromy matrix is given by a commutator,
\begin{equation}\label{eq:BoundMonToTime}
\partial_\tau T_b (z) = \left[ T_b (z) , N(z) \right] \, ,
\end{equation}
for some matrix $N(z)$ one will indeed find that  $\partial_\tau \Tr T_b(z)^{n} =0$ for any $n\in \mathbb{N}$ and $z\in\mathbb{C}$. Assuming\footnote{In general  this strongly depends on the specific form of the Lax connection $\mathcal{L}(z)$
but the  procedure described here can be easily adapted to other cases. },
 \begin{equation}\label{eq:ReflectedTransport}
 T^\Omega_R (2\pi , \pi ;z) = T^\Omega(0,\pi ; z_R)\, ,  
 \end{equation}
we find  explicitly using \eqref{eq:TransportToTime} that the time derivative of the boundary monodromy matrix satisfies,
\begin{equation}
\begin{aligned}
\partial_\tau T_b (z) =\; & \left[ T^\Omega(0,\pi; z_R) \Omega[\mathcal{L}_\tau ( \tau, \pi; z_R)] - \Omega[\mathcal{L}_\tau ( \tau , 0 ;z_R)] T^\Omega(0,\pi ; z_R) \right]  T(\pi , 0 ; z)  \\
& + T^\Omega(0,\pi ; z_R)  \left[ T(\pi ,0 ; z)\mathcal{L}_\tau ( \tau , 0 ;z) - \mathcal{L}_\tau ( \tau, \pi ; z) T(\pi,0;z) \right] \, .
\end{aligned}
\end{equation}
The  condition \eqref{eq:BoundMonToTime}  sufficiently holds when $N(z) =\mathcal{L}_\tau (\tau , 0;z)$ and when we require the following boundary conditions on both the endpoints\footnote{In principle the boundary conditions can be different on each endpoint (see e.g.\ \cite{Driezen:2018glg}) which in the string theory application allows the open string to connect distinct D-brane configurations.},
\begin{equation}\label{eq:BoundCondLax1}
\mathcal{L}_\tau (\tau, 0 ;z ) = \Omega[ \mathcal{L}_\tau (\tau ,0 ; z_R) ] \, ,
\end{equation}
and similarly on $\sigma = \pi$. When studying a specific two-dimensional integrable model with a known Lax connection, and knowing its behaviour under the reflection $R$, one can now easily derive the \textit{integrable} boundary conditions on the field variables by eq.~\eqref{eq:BoundCondLax1}.  Typically this will involve additional conditions on the automorphism $\Omega$ as we will see in the coming section.

\section{Applied to $\lambda$-deformations} \label{s:lambda}

We will now apply the boundary monodromy method to the (standard) $\lambda$-deformation introduced by Sfetsos in \cite{Sfetsos:2013wia}. The interest in this particular model is that it is a two-dimensional integrable field theory deforming the exactly conformal Wess-Zumino-Witten (WZW) model on group manifolds.  We will therefore be able to relate the integrable boundary conditions of the $\lambda$-model to known results of stable D-brane configurations  wrapping twisted conjugacy classes in the group manifold \cite{Alekseev:1998mc,Felder:1999ka,Stanciu:2000fz,Figueroa-OFarrill:2000lcd,Bachas:2000fr}.

\subsection{Construction of the $\lambda$-action}

Let us first briefly introduce the construction of  $\lambda$-deformations  of \cite{Sfetsos:2013wia}. One starts by doubling the degrees of freedom on a Lie group manifold $G$, by combining the WZW model on $G$ at level $k$ with the Principal Chiral Model (PCM)  on $G$ with a coupling constant $\kappa^2$, i.e.,
 \begin{equation}\label{eq:doubledaction}
 \begin{aligned}
 S_{k,\kappa^2}(g,\widetilde{g}) &= S_{\text{WZW},k}(g) + S_{\text{PCM},\kappa^2}(\widetilde{g}) ,\\
   S_{\text{WZW,k}}(g) &= -\frac{k}{2\pi}\int_\Sigma  d  \sigma d\tau    \langle g^{-1} \partial_+ g , g^{-1} \partial_- g \rangle - \frac{  k}{24\pi }       \int_{M_3}   \langle   \bar g^{-1} d\bar g, [\bar g^{-1} d\bar g,\bar g^{-1} d\bar g]  \rangle , \\  
 S_{\text{PCM},\kappa^2}(\widetilde{g}) &= -  \frac{\kappa^2}{\pi} \int d\sigma d\tau \, \langle \widetilde{g}^{-1}\partial_+ \widetilde{g} , \widetilde{g}^{-1}\partial_- \widetilde{g} \rangle \, , 
 \end{aligned}
 \end{equation}
which are both realised through distinct group elements $g\in G$ and $\widetilde{g}\in G$ respectively\footnote{We have taken    conventions on the worldsheet $\Sigma$  in which the two-dimensional metric is  fixed  as $\text{diag}(+1,-1)$, $\epsilon_{\tau\sigma} = 1$ and $\partial_\pm = \frac{1}{2}(\partial_\tau \pm \partial_\sigma)$. Moreover, we consider compact semi-simple Lie groups $G$ of which the generators $T_A$, $A \in \{1, \cdots, \text{dim}(G) \}$ of the Lie algebra $\mathfrak{g}$ are Hermitean and normalised with respect to the  Cartan-Killing billinear form $\langle \cdot , \cdot \rangle$ as $\langle T_A, T_B \rangle = \frac{1}{x_R} \Tr (T_A T_B) = \delta_{AB}$ (with $x_R$ the index of the representation $R$). It is known that for compact groups the level $k$ should be integer quantised while for non-compact groups it can be free \cite{Witten:1983ar}.}. The fields $\bar{g}$ are an extension of $g$ into $M_3 \subset G$ such that $\partial M_3 = g(\Sigma)$. Altogether the doubled model  \eqref{eq:doubledaction} has a global $G_L \times G_R$ symmetry. Next, one reduces back to $\text{dim}(G)$ degrees of freedom by gauging a subgroup acting as,
\begin{equation}
\begin{aligned}
G_L : \widetilde{g} \rightarrow h^{-1} \widetilde{g}, \qquad G_{\text{diag}}: g \rightarrow h^{-1} g h, \qquad \text{with} \; h \in G,
\end{aligned}
\end{equation}
using a common gauge field $A \rightarrow h^{-1} A h - h^{-1} d h$. Doing a minimal substitution on the PCM, by replacing $\partial_\pm \widetilde{g} \rightarrow D_\pm \widetilde{g} = \partial_\pm \widetilde{g} - A_\pm \widetilde{g}$ and replacing the WZW model by the $G/G$ gauged WZW model, 
  \begin{equation}
  S_{\text{gWZW,k}}(g,A) = S_{\text{WZW},k}(g) + \frac{k}{\pi} \int d\sigma d\tau \, \langle A_- , \partial_+ g g^{-1} - A_+ ,  g^{-1}\partial_- g  + A_+ , g^{-1} A_- g  -  A_+ , A_- \rangle \ .
  \end{equation}
one finds the $\lambda$-deformation after fixing the gauge by $\widetilde{g} = \mathbf{1}$,
\begin{equation}
  \begin{aligned}\label{eq:LambdaAction1}
  S_{k,\lambda}(g, A) =  S_{\text{WZW},k} (g) 
  - \frac{k}{ \pi} \int d\sigma d\tau \langle A_+ ,  (\lambda^{-1} - D_{g^{-1}}) A_- \rangle
 - \langle A_- , \partial_+ g g^{-1} \rangle + \langle A_+ , g^{-1}\partial_- g   \rangle   \ .
  \end{aligned}
  \end{equation}
Here we have introduced the adjoint operator $D_g : \mathfrak{g} \rightarrow \mathfrak{g}$, $D_g (T_A) = g T_A g^{-1} = T_B (D_g){}^B{}_A$ with $g\in G$ and the parameter $\lambda$,
\begin{equation}
\lambda = \frac{k}{k+\kappa^2} .
\end{equation}
The gauge fields are now auxiliary and can be integrated out. Varying  the action $S_{k,\lambda}(g,A) $ with respect to $A_{\pm}$ we find the constraints,
\begin{equation}\label{eq:GaugeConstraints}
A_+ = \left( \lambda^{-1} -  D_g \right)^{-1} \partial_+ g g^{-1}\, , \qquad   
A_-  = -\left( \lambda^{-1} - D_{g^{-1}} \right)^{-1} g^{-1} \partial_- g\, .
\end{equation}  
Substituting these  into eq.~\eqref{eq:LambdaAction1} gives the large $k$ effective action,
\begin{equation}\label{eq:LambdaAction2}
\begin{aligned}
S_{k,\lambda}(g) = S_{\text{WZW},k}(g) - \frac{k }{\pi}\int \mathrm{d}\sigma \mathrm{d}\tau \, \partial_+ g g^{-1} \left( \lambda^{-1} -  D_{g^{-1}} \right)^{-1} g^{-1}\partial_- g \ , 
\end{aligned}
\end{equation}
which is an all-loop in $\lambda$ deformation of the WZW theory with a global $g \rightarrow g_0 g g_0^{-1}$, $g_0\in G$ symmetry left.  Effectively, the $\lambda$-theory thus  deforms the target space metric and Kalb-Ramond field  of the WZW $\sigma$-model. In addition, the Gaussian elimination of the gauge fields in the path integral results in a non-constant dilaton profile,
\begin{equation}
\Phi = \Phi_0 - \frac{1}{2} \ln \det \left( \mathbf{1} - \lambda D_{g^{-1}} \right) , 
\end{equation}
with $\Phi_0$ constant. 
While the integrability of the $\lambda$-model (with periodic boundary conditions) has been first shown in \cite{Sfetsos:2013wia} starting from the effective $\sigma$-model action \eqref{eq:LambdaAction2}, one can straightforwardly show it starting from \eqref{eq:LambdaAction1} as  done in  \cite{Hollowood:2014rla}. The Lax connection ${\cal L}(z)$ representing the equations of motion of the fields $g$ satisfying the zero-curvature condition \eqref{eq:zerocurvatureLax} $\forall\ z \in \mathbb{C}$ is,
\begin{equation} \label{eq:LambdaLax}
{\cal L}_\pm (z) = - \frac{2}{1 +\lambda} \frac{A_\pm}{1 \mp z} ,
\end{equation}
upon the constraints \eqref{eq:GaugeConstraints}. 

\subsection{Interpretation as (twisted) conjugacy classes}
To apply the boundary monodromy method to the $\lambda$-model  we first need to consider the behaviour of the transport matrix $T_R^\Omega (2\pi, \pi ; z)$ under the reflection $R: \sigma \rightarrow 2\pi - \sigma$. For the $\lambda$-Lax \eqref{eq:LambdaLax} one will find that eq.\  \eqref{eq:ReflectedTransport} holds for $z_R = -z$. The resulting integrable boundary conditions \eqref{eq:BoundCondLax1} of the $\lambda$-model are then, after expanding order by order in the spectral parameter $z$,
\begin{equation}
\left. A_+ \right\vert_{\partial\Sigma} =\left.  \Omega \left[ A_- \right] \right\vert_{\partial\Sigma},
\end{equation}
together with the requirement that $\Omega$ is an \textit{involutive} automorphism of the Lie algebra, 
\begin{equation}
\Omega^2 = 1 .
\end{equation}
Moreover, to interpret the above boundary conditions as Dirichlet and (generalised) Neumann conditions, the automorphism $\Omega$ should be such that no energy-momentum is flowing through the boundary, i.e. the energy-momentum tensor must satisfy $T_{01}| = 0$, which  turns out to require that it is metric-preserving in the sense of $\langle \Omega (T_A) , \Omega (T_B) \rangle = \langle T_A , T_B \rangle$.

Upon the constraint \eqref{eq:GaugeConstraints} the integrable boundary conditions are now given by,
\begin{equation} \label{eq:intbc}
(\mathbf{1} - \lambda D_g)^{-1} \partial_+ g g^{-1} = - \Omega (\mathbf{1} - \lambda D_{g^{-1}})^{-1} g^{-1}\partial_- g .
\end{equation}
At the WZW conformal point ($\lambda = 0$) one  consistently finds the (twisted) gluing conditions of \cite{Alekseev:1998mc,Felder:1999ka,Stanciu:2000fz} of the holomorphic Kac-Moody currents $J_+ = -k\partial_+ g g^{-1},\, J_- = kg^{-1} \partial_- g$  on the boundary, 
\begin{equation} \label{eq:wzwgluing}
\lambda \rightarrow 0: \qquad    J_+ = \Omega (J_- ) ,
\end{equation}
preserving precisely one copy of both the Kac-Moody current algebra (iff.\ $\Omega$ is a Lie algebra automorphism, which is the case here) and the Virasoro algebra. Because only the latter property is a necessity to preserve conformal invariance, the former property led to the description of  the corresponding  D-brane configurations as being `maximally symmetric'.  In \cite{Alekseev:1998mc,Felder:1999ka,Stanciu:2000fz} (see also \cite{Figueroa-OFarrill:1999cmq}) it was  shown, starting from the corresponding Dirichlet conditions of eq.~\eqref{eq:wzwgluing},  that the D-brane worldvolumes  wrap (twisted) conjugacy classes of the group $G$,
\begin{equation}
C_\omega (g) = \{ h g \omega (h^{-1} ) \, | \, \forall h\in G \} , \qquad \omega (e^{tX}) \equiv e^{t \Omega (X)} \in G , \;\; X\in \mathfrak{g},
\end{equation}
classified by the quotient of metric-preserving outer automorphisms $\omega \in \text{Out}_0 (G) = \text{Aut}_0(G)/\text{Inn}_0(G)$ \cite{Figueroa-OFarrill:1999cmq}. When $\omega \in \text{Inn}_0(G)$ is inner, i.e. $\omega(h) = \text{ad}_w (h) = w h w^{-1}$ for some $w\in G$, the twisted conjugacy class $C_\omega (g)$ is related to the ordinary conjugacy class $C_{\text{Id}}(g)$ by a (right) group translation,
\begin{equation}
C_{\text{ad}_w} (g) = C_{\text{Id}} (gw) w^{-1} ,
\end{equation}
which is a symmetry of the WZW model. The automorphisms $\omega$ are in principle not constrained any further here.

For generic $\lambda$ it was shown in \cite{Driezen:2018glg} that the geometrical picture of the integrable boundary conditions \eqref{eq:intbc}  as D-branes wrapping twisted conjugacy classes persists by pleasing cancellations of the $\lambda$-dependence. This is indeed expected, since the deformation affects only  target space data  as the metric, while the worldvolumes are defined through the orthogonal decomposition of the tangent space with respect to the Dirichlet conditions, independently of the target space metric.  However, integrability picks out only the automorphisms $\omega (e^{tX}) = e^{t\Omega(X)}$  that satisfy $\Omega^2 = 1$.  Generic inner automorphisms, or group translations of the conjugacy classes, are thus excluded. Indeed, independent right group translations are not a symmetry of the $\lambda$-model which remarkably follows from demanding integrability structures.

\subsection{$G = SU(2)$ illustration}

\begin{figure}[H]
\centering
\includegraphics[scale=0.15]{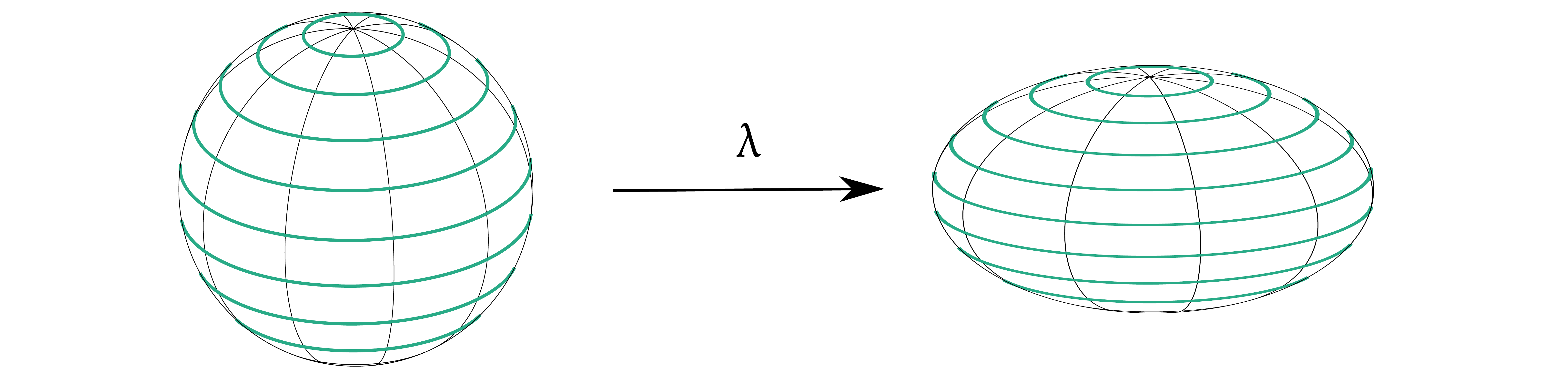}
\caption{For illustrative purposes we portray here the $S^3 \simeq SU(2)$ group manifold in one dimension less. The green lines represent the $S^2$-branes or conjugacy classes in the $S^3$ that change size under the \textit{squashing} of the $S^3$ when the $\lambda$-deformation is turned on.}\label{fig:su2illustration}
\end{figure}

To illustrate the above, we focus in this section on the case of the  $G= SU(2)$ group manifold for which $\text{Out}_0(SU(2)) = \text{Id}$ is trivial and one will describe ordinary conjugacy classes\footnote{ For an analysis and explicit example of \textit{twisted} conjugacy classes, possible in $\lambda-SL(2,R)$, we refer to \cite{Driezen:2018glg}. Interestingly, in $G=SL(2,R)$ only the twisted conjugacy classes turn out to correspond to \textit{stable} D-brane configurations \cite{Bachas:2000fr}, telling us it is indeed worth including the possibility of twisted gluing conditions.}. We parametrise the group element $g\in SU(2)$ in Cartesian coordinates,
\begin{equation}
g = \begin{pmatrix}
X_0 + i X_3 & - X_1 + i X_2 \\ X_1 +  i X_2 & X_0 - i X_3 
 \end{pmatrix} ,
\end{equation}
constrained to $\det g = X_0^2 + X_1^2 + X_2^2 + X_3^2 = 1$,  making the embedding of $SU(2)$ as an $S^3$ in $\mathbb{R}^4$ apparent. The set of group elements in an ordinary conjugacy class $C(g)$ clearly have a constant trace and here  fix the $X_0$ parameter  to some constant value. The conjugacy classes are thus $S^2$-spheres of varying radius inside the $S^3$ as illustrated  in figure \ref{fig:su2illustration}. When the $\lambda$-parameter is turned on, and the target space metric $G$ gets deformed, what is changed will be the size (or radius) of the $S^2$-spheres by the induced deformed metric $\left. G\right\vert_{S^2}$  on these branes. In figure \ref{fig:su2illustration} it is moreover clear that  the rotational symmetry of the WZW gets lost in the deformation, which is a reflection of the $\Omega^2 = 1$ constraint coming from integrability and preventing the existence of  rotated branes. Indeed, in \cite{Driezen:2018glg} a semi-classical analysis of the quadratic scalar fluctuations of the branes moreover shows that at $\lambda=0$ a massless  p-wave triplet exists which becomes massive for $\lambda \neq 0$.
In both cases, there is a total of 2 static D0-branes (corresponding to the north- and southpole) and $l$ static D2-branes (corresponding to the $S^2$'s). The number $l$ is integer quantised and equal to $l=k-1$, following from topological obstructions in formulating the WZ term in \eqref{eq:doubledaction} in the presence of a boundary. In the boundary case  the WZ term should be altered as \cite{Klimcik:1996hp},
\begin{equation}\label{eq:wzbdy}
\int_{M_3} H \rightarrow \int_{M_3} H + \int_{D_2}
 \omega ,
 \end{equation}
 with $\partial M_3 = g(\Sigma) + D_2$, $D_2 \subset g(\partial\Sigma)$ and $\omega$ a two-form on $D_2$ such that $H |_{D_2} = \mathrm{d}\omega$. To cancel  in the path integral  ambiguities in the choice of $M_3$ recall that the closed string WZW theory on compact groups requires the level $k$ to be integer quantised. On the other hand, the open string WZW theory with \eqref{eq:wzbdy} will require the D-branes to be localised on a discrete number of positions. In the case of $G=SU(2)$ the number of branes are then indeed fixed as $l = k-1 \in \mathbb{Z}$, where we refer to \cite{Klimcik:1996hp,Stanciu:2000fz,Figueroa-OFarrill:2000lcd} for more details. Interestingly, this can be seen as a semi-classical stabilisation of the D2-branes, since their localised positions forbids them to shrink smoothly to zero size. When the deformation is turned on, \cite{Driezen:2018glg} shows the \textit{continuous} $\lambda$-dependence precisely cancels in the topological conditions\footnote{Both the $H$-form and $\omega$-form receive a $\lambda$-contribution but these precisely cancel.} and so consistently also in the $U(1)$ flux quantisation. Again, indeed, a semi-classical analysis of the scalar fluctuations gives a massive s-wave with a mass independent of $\lambda$.

\subsection{Interplay with generalised T-dualities}
Another motivation to look at $\lambda$-deformations is the close connection to generalised T-dualities. The $\lambda \rightarrow 1$ scaling limit (obtained by taking $k \rightarrow \infty$) produces e.g.\ the non-Abelian T-dual of the Principal Chiral Model (PCM) \cite{Sfetsos:2013wia}. On the other hand, for generic values of $\lambda \in [0,1]$ the model is Poisson-Lie T-dual \cite{Klimcik:1995ux,Klimcik:1995dy}, up to an additional analytical continuation, to an integrable deformation of the PCM  \cite{Vicedo:2015pna,Hoare:2015gda,Klimcik:2015gba}  known as Yang-Baxter or $\eta$-deformations \cite{Klimcik:2008eq,Delduc:2013fga} which have an action,
\begin{equation}
S_{t, \eta} (\widehat{g}) = \frac{1}{t} \int_\Sigma d\sigma d\tau\; \partial_+ \widehat{g} \widehat{g}^{-1} \left( \mathbf{1} - \eta {\cal R} \right)^{-1} \partial_- \widehat{g} \widehat{g}^{-1} ,
\end{equation}
where ${\cal R} : \mathfrak{g} \rightarrow \mathfrak{g}$ is an operator solving the modified classical  Yang-Baxter equation. 

 Tracking the behaviour of D-brane configurations under these generalised T-dualities is, in general, a challenging procedure due to the lack of well-defined boundary conditions in the curved background geometry. In the $\lambda$-deformation, however, integrability dictates us precise boundary conditions (in eq.~\eqref{eq:intbc}) given   in terms of worldsheet derivatives of the phase-space variables. Together with the known canonical transformations of non-Abelian T-duality \cite{Lozano:1995jx,Sfetsos:1996pm} (NATD) and Poisson-Lie (PL) T-duality \cite{Sfetsos:1997pi} this allows us to find the dual D-branes in the e.g.  $G= SU(2)$ case. Schematically we find  \cite{Driezen:2018glg},
 \begin{equation*}
\begin{aligned} 
&\text{D2-brane in the NATD of the PCM }  \quad &&\xrightarrow{\text{can.\ transf.\ }} \quad  &&\text{D3-brane in the original PCM} \\
 &\text{D2-brane in the $\lambda$-deformed WZW} \quad &&\xrightarrow{\text{can.\ transf.\ }} \quad &&\text{D3-brane in the $\eta$-deformed PCM}
 \end{aligned} 
 \end{equation*}
 so that in both cases the $S^2$-branes pop open to space-filling D3-branes. Remarkably, the boundary conditions obtained in this way in \cite{Driezen:2018glg} match precisely with the boundary conditions that follow from the boundary monodromy method in section \ref{s:bmm} when plugging in the Lax pair of the Principal Chiral Model \cite{Zakharov:1973pp},
 \begin{equation}
 {\cal L }_\pm (z) = \frac{1}{1\mp z} g^{-1} \partial_\pm g,
 \end{equation}
 or of the $\eta$-deformed PCM \cite{Klimcik:2008eq,Delduc:2013fga},
 \begin{equation}
 {\cal L }_\pm (\eta ; z) = \frac{1+\eta^2}{1\pm z} D_g \cdot \frac{1}{1\pm \eta {\cal R}} \cdot \partial_\pm g g^{-1} ,
 \end{equation}
in \eqref{eq:BoundCondLax1} respectively. 


\section{Conclusions and outlook}\label{s:concl}
In this overview note we have seen an efficient method to derive classical integrable boundary conditions in $\sigma$-models by demanding that the monodromy matrix of the Lax connection generates conserved charges even in the presence of boundaries. As emphasized in the introduction, this is generically challenging for stringy $\sigma$-models without precise CFT formulations. The boundary monodromy method, however, essentially  only requires  the knowledge of the $\sigma$-model Lax connection. \\
\indent In the context of  $\lambda$-deformations the boundary monodromy method dictates us integrable boundary conditions that are described elegantly by twisted conjugacy classes in the deformed target space. This geometrical picture is independent of the deformation parameter and, indeed, corresponds smoothly to the D-brane configurations dictated by CFT methods of the undeformed WZW point. Illustrated in the $SU(2)$ manifold we have seen that the deformation changes the size of the D-branes and destroys their rotational symmetry in the deformed geometry.  This latter (natural) observation ties nicely together with the constraining features of integrability. Additionally, we have seen that the flux quantisation consistently remains independent of the $\lambda$-parameter and enforces the branes to sit stabilised at localised positions. These conclusions are supported in \cite{Driezen:2018glg} by an analysis of  scalar fluctuations of the D-branes. Finally, armed with precise integrable boundary conditions, one can track them under generalised dualities present in  $\lambda$-deformations. For $G=SU(2)$ we have seen a Dirichlet condition to transform into a generalised Neumann which, to close the circle, turns out to follow from demanding integrability of the dual models as well.\\

Let us stress the analysis so far has been purely classical. It remains an interesting question to understand the quantum description of the integrable boundary conditions in these $\lambda$-models. Here, the bulk $S$-matrix of \cite{Appadu:2017fff} should be supplemented by a boundary $K$-matrix describing particle reflections off the boundary and satisfying the boundary Yang-Baxter equation \cite{Cherednik:1985vs,Sklyanin:1988yz}. Since the $S$-matrix of \cite{Appadu:2017fff} was derived by mapping the quantum $\lambda$-model to  a spin $k$ XXX Heisenberg spin chain, it would be appealing to interpret the boundary $K$-matrix in the corresponding open spin chain as well.\\
\indent Another appealing line of study, returning to the string point of view, is to consider integrable D-branes of $\lambda$-models in supercoset geometries \cite{Hollowood:2014qma} as here the deformation is expected to be truly marginal to all loops.

\section*{Acknowledgements}
I would like to thank the organisers of the ``Dualities and Generalised Geometries" session part of the Corfu Summer Institute 2018 schools and workshops for giving me the opportunity to speak and for a stimulating and interesting workshop. Additionally, I would like to thank my collaborators and advisers Daniel Thompson and Alexander Sevrin  for  the fruitful discussions we have had in the process of \cite{Driezen:2018glg} and Saskia Demulder for a careful read of the manuscript. Finally, I acknowledge also the  support by the ``FWO-Vlaanderen'' through an aspirant fellowship and the project  G006119N.

%

\bibliographystyle{/Users/sibylledriezen/Dropbox/PhD:Bibfile/JHEP}
\bibliography{/Users/sibylledriezen/Dropbox/PhD:Bibfile/SibBib}

\end{document}